\documentclass[useAMS,usenatbib]{mn2e}
\usepackage{graphicx}
\usepackage{psfig}
  
\def\kms{\relax \ifmmode {\,\rm km\,s}^{-1}\else \,km\,s$^{-1}$\fi}

\def\mincir{\ \raise-2.truept\hbox{\rlap{\hbox{$\sim$}}\raise5.truept
    \hbox{$<$}\ }}
\def\magcir{\ \raise-2.truept\hbox{\rlap{\hbox{$\sim$}}\raise5.truept
    \hbox{$>$}\ }}
\def\gr{$^\circ$}
\def\sm{M$_\odot$}

\def\arcsec{\hbox{$^{\prime\prime}$}}
\def\arcmin{\hbox{$^{\prime}$}}

\def\hi{H~{\sc i}}
\def\hii{H {\sc ii}}

\def\oiii{[O {\sc iii}]}

\title[IC10: PN kinematics]{A kinematic study of planetary nebulae 
in the dwarf irregular galaxy IC10
\thanks{Based on data collected at the Subaru Telescope, which is operated by
    the National Astronomical Observatory of Japan.}}

\author[Gon\c calves, et al.]
  {Denise R. Gon\c calves$^{1,2}$\thanks{E-mail: denise@astro.ufrj.br}, 
Ana M. Teodorescu$^{3}$, 
  Alan Alves-Brito$^{4}$,
  \newauthor 
  Roberto H. M\'endez$^{3}$ and Laura Magrini$^{5}$ 
\\
  $^{1}$ Observat\'orio do Valongo - Universidade Federal do Rio 
 de Janeiro, Ladeira Pedro Antonio 43,
 20080-090 Rio de Janeiro, Brazil\\
  $^{2}$ Department of Physics and Astronomy, University College London,
 Gower Street, WC1E 6BT  London, UK\\
  $^{3}$ Institute for Astronomy, University of Hawaii, 
2680 Woodlawn Drive, HI 96822 Honolulu, USA\\
  $^{4}$ Research School of Astronomy and Astrophysics,
The Australian National University, Cotter Road, Weston, ACT 2611, 
Australia\\
  $^{5}$ INAF - Osservatorio Astrofisico di Arcetri, 
Largo E. Fermi 5, I-50125 Firenze, Italy\\
} 

\begin{document}



\maketitle

\label{firstpage}

\begin{abstract}
We present positions, kinematics, and the planetary nebula luminosity
function (PNLF) for 35 planetary
nebulae (PNe) in the nearest starburst galaxy IC10 extending out to 
3~kpc from the galaxy's centre. We take advantage of the deep imaging and 
spectroscopic capabilities provided by the spectrograph FOCAS on 
the 8.2m Subaru telescope. The PN velocities were measured
through the slitless-spectroscopy technique, which allows us to explore the
kinematics of IC10 with high precision. Using these velocities, we conclude
that there is a kinematic connection between the \hi\ envelope located around
IC10 and the galaxy's PN population. By assuming that the PNe in the central 
regions and in the outskirts have similar ages, our results put strong 
observational constraints on the past tidal interactions in the Local Group.
This is so because by dating the PN central stars, we, therefore, infer the
epoch of a major episode of star formation likely linked to the
first encounter of the \hi\ extended envelope with the galaxy. 
 Our deep \oiii\ images also allow us to use the 
PNLF to estimate a distance modulus of 24.1$\pm0.25$, which is in agreement 
with recent results in the literature based on other techniques.

\end{abstract}


\section[]{Introduction}
Even though dwarf galaxies are currently the most 
abundant galaxy population, and were 
probably even more abundant during the first epochs of the Universe,  
the theory of their formation is still under debate  \citep{mateo98}. 
According to the hierarchical formation scenario 
\citep[see, e.g.,][]{white91}, they are potential building blocks of
larger structures, so their study is crucial 
for understanding galaxy formation and evolution processes. 
One of the several questions about 
their formation and evolution, and that still remains unsolved, is the origin
of the differences between the irregular and spheroidal morphological types
\citep[see][for a review]{mateo98}.
We have been approaching this issue by using planetary nebulae (PNe,
old-intermediate stellar population) and \hii\ regions (young population) as
tracers of the chemical evolution of the Local Group (LG) 
dwarf galaxies \citep[see, for example,][]{gon07,mg09,m11,gon12}. 
In particular, we have dedicated the last few years in the  study of 
the dwarf irregular galaxy IC10 -- its PNe
and \hii\ regions \citep{magrini03,mg09}; its serendipitously discovered
symbiotic system, which is the farthest known symbiotic star to date
\citep{gon08}; and its detailed chemical
evolution history \citep{yin10}.  
Here we present a kinematic study of PNe in this galaxy. As a by-product, 
we also present some photometric results.

IC10 is a dwarf irregular galaxy located  in the outskirts of the 
Local Group at a low galactic latitude (l = 119$^{o}$.0, b = --3$^{o}$.3;
NASA/IPAC Extragalactic Database, NED), which makes the reddening values very
uncertain. The estimates of the
reddening range from E(B-V)=0.47 to 2.0, translating into a distance modulus
($m-M$) that varies from 22 to more than 27 mag, i.e., the linear distance for
IC 10 would vary from 0.5 to 3~Mpc (Sakai, Madore  \& Freedman 1999; Demers, 
Battinelli \& Letarte 2004). More recent estimates locate IC10 at a
distance  around 0.7-0.8 Mpc (e.g., Pustilnik \& Zucker 2008; 
\citealt{sanna08}; \citealt{kim09}), suggesting a possible
membership to the M31 subgroup.

In addition, IC10 has a high star formation
rate (SFR) as evidenced by its large number of \hii\ regions 
\citep{hl90} as well as for its H$\alpha$ \citep{mateo98} and far-IR 
 \citep{mi94} luminosity. It also has a huge number of Wolf-Rayet
stars per unit luminosity, the largest in the LG (Massey, Armandroff 
\& Conti 1992; \citealt{ma95}), and an anomalous ratio of
carbon-type Wolf-Rayet stars (WC stars) to nitrogen-type Wolf-Rayet
stars (WN stars), which is peculiar at its metallicity (0.2--0.3 solar;
cf. \citealt{skh89} and \citealt{garnett90}). The presence
of  an extended and complex \hi\ envelope \citep[][hereafter SS89]{shostak89}
suggests that  IC10 is still in its formative
stage or that it has undergone a major merger episode \citep[][hereafter
WM98]{wilcots98}. Altogether these characteristics suggest 
that IC10 is experiencing
an intense and very recent burst of star formation, likely
triggered by infalling gas from its \hi\ envelope 
which is counter-rotating with 
respect to the galaxy proper rotation (WM98).

The Local Group Census (cf. \citealt{cm06}) 
discovered a large number of planetary nebulae (PNe) 
in many dwarf galaxies, including IC10.  
Several of these PNe were found at large distances from
the centre of their host galaxies, 
and so could represent a trace of past tidal interactions and/or
a connection of the old stellar population 
with the \hi\ envelope located around some dwarf 
galaxies, as in the case of IC10 \citep{huctmeier79}.  

Single-dish 21~cm observations showed that IC10 
is embedded within a huge envelope of \hi\ 
extending seven times farther than its optical 
diameter and stretching nearly a degree on the
sky (\citealt{huctmeier79}; Huchtmeier, 
Seiradakis \& Materne 1981). Moreover, the \hi\ velocity field
of IC10 appears to be quite complex \citep{huctmeier79} with a long
stream extending to the South of the galaxy. 
Interferometric data revealed that IC10 has a
rotating disk surrounded by a counter-rotating 
or highly warped outer distribution
of gas (SS89).

The most likely scenario is that the gas 
component represents a late acquisition.
Numerical simulations, for example, indicate 
that mergers can lead to counter-rotating
cores (e.g., \citealt{thakar97}).  The 
detection of a stellar population
related to the outer envelope would 
be extremely important to constrain its age and thus the
evolution of IC10, dating the possible encounter between the two components.  
However, so far deep
optical images did not show any stellar 
population coincident with the extended gas
envelope, as discussed by WM98 from their 
examination  of Digital Sky Survey and deep WIYN R band images. 
The surface brightness of this stellar 
population is expected to be very low, and
its detection is strongly limited by scattered 
light from nearby bright objects and by the
contribution of faint discrete sources. These 
limitations can be circumvented
observing PNe because they are easily detectable 
from their strong nebular emission of \oiii\ at 5007~\AA.  

In this paper we aim at studying the radial velocities of the PN population 
of IC10 in order to explore the connection of its intermediate to old 
stellar population with its \hi\ envelope. Previous to the present work, 
the PN 
kinematics of this galaxy was never explored, although a number of PNe were 
already known. 
Altogether sixteen candidate PNe were identified in IC10 by the Local Group 
Census in 
\citet{magrini03}. Other nine fainter PNe were detected and immediately 
confirmed using deep spectroscopy by \citet{mg09}.

\section[]{Data acquisition and reduction} 

The observations were made by one of us (R.H.M.) with the
Faint Object Camera and Spectrograph 
(FOCAS; \citealt{kashikawa02}) attached to the 
Cassegrain focus of the 8.2 m Subaru telescope, Mauna Kea, Hawaii.

\subsection[]{Observations}

Data were obtained during a single  night, 2010 September 30th.
The night was photometric, with an average seeing of 0$''$.8.

The field of view of FOCAS is 6.5 arcmin and is covered by
two CCDs of 2k $\times$ 4k (pixel size 15 $\mu$m) with an unexposed
gap of 5$''$ between them. For simplicity, we will call the two CCDs
Chip 1 and Chip 2. The image scale is 0.104 arcsec pixel$^{-1}$.

\begin{figure} 
   \centering
   \includegraphics[width=8 truecm]{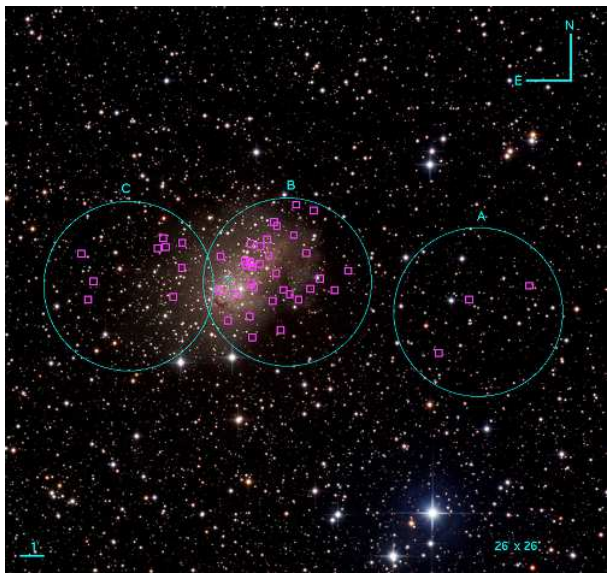} 
   \caption{The distribution of all PN candidates (squares) detected with 
     Subaru+FOCAS in IC10. The big circles indicate the
     6.5~arcmin FOCAS field of view. 
At the distance of 660~kpc (see section 4.2), 
     one arcmin corresponds to 0.19~kpc.}
   \label{fig_FocasFields}
\end{figure}

The observations were obtained with two filters: the on-band filter with 
a central wavelength of 5025~\AA, a FWHM (Full Width Half Maximum) of 60~\AA, a
peak transmission of 68\%, and an equivalent width of 40~\AA, and the
off-band filter image was done through a standard broadband visual filter.
We observed three fields -- A, B, and C -- centred respectively at
($\alpha$,$\delta$, J2000) = (0h 19m 00.4s, 59\gr16\arcmin25\arcsec; 0h 20m
03.9s, 59\gr17\arcmin18\arcsec; 0h 20m 56.9s, 59\gr16\arcmin48\arcsec), as
shown in Figure~\ref{fig_FocasFields}. The observing sequence 
for each field was: one off-band image (exposure time 150 s), 
one on-band image (exposure time 1200 s) and two grism$+$on-band 
images (exposure time 1200 s). This observing sequence was executed
twice for fields A and B and three times for field C.

For the wavelength calibration, after each grism$+$on-band exposure, 
an engineering mask was inserted in
the light path, and on-band and grism$+$on-band images were
obtained illuminating the mask with the comparison lamp. In
addition, on-band and grism$+$on-band images of the engineering
mask were taken illuminating the mask with the Galactic
PN NGC 7293 (PNG 036.1-57.1), for radial velocity quality
control. Examples of the calibration images are shown in Figure~2 
of \citet{mendez09}. Table~1 shows the log 
of the observations with the corresponding exposure 
times and air masses. For brevity, we 
list only the images corresponding to Chip 1. 
Identification numbers of Chip 2 images are 
obtained by adding 1 to the Chip 1 identification numbers in Table 1.
The spectrophotometric standard G 138-31 
\citep{oke90} was used for the photometric calibration of the on-band 
images.

\begin{table}
\centering
\begin{minipage}{80mm}
\caption{Journal of observations for Chip 1.}
\scriptsize{
\begin{tabular}{@{}lllrl@{}}
\hline
\ FOCAS Field &         Configuration & FOCAS id & Exp & Air Mass\\       
\hline
 G 138 - 31              &   On-band  &  120371  &   60  &  1.33 \\
 G 138 - 31              &   On-band  &  120373  &   60  &  1.36 \\
 NGC 7293 + mask         &   On-band  &  120397  &  300  &  1.87 \\
 NGC 7293 + mask         &  On+grism  &  120399  &  600  &  1.80 \\
 IC 10 Field B           &  Off-band  &  120403  &  150  &  1.94 \\
 IC 10 Field B           &   On-band  &  120405  & 1200  &  1.86 \\
 IC 10 Field B           &  On+grism  &  120407  & 1200  &  1.75 \\
 IC 10 Field B           &  On+grism  &  120409  & 1200  &  1.65 \\
 IC 10 Field B           &  Off-band  &  120423  &  150  &  1.46 \\
 IC 10 Field B           &   On-band  &  120425  & 1200  &  1.43 \\
 IC 10 Field B           &  On+grism  &  120427  & 1200  &  1.39 \\
 IC 10 Field B           &  On+grism  &  120429  & 1200  &  1.36 \\
 IC 10 Field A           &  Off-band  &  120431  &  150  &  1.33 \\
 IC 10 Field A           &   On-band  &  120433  & 1200  &  1.32 \\
 IC 10 Field A           &  On+grism  &  120435  & 1200  &  1.31 \\
 IC 10 Field A           &  On+grism  &  120437  & 1200  &  1.30 \\
 IC 10 Field C           &  Off-band  &  120443  &  150  &  1.30 \\
 IC 10 Field C           &   On-band  &  120445  & 1200  &  1.30 \\
 IC 10 Field C           &  On+grism  &  120447  & 1200  &  1.31 \\
 IC 10 Field C           &  On+grism  &  120449  & 1200  &  1.32 \\
 IC 10 Field C           &  Off-band  &  120451  &  150  &  1.33 \\
 IC 10 Field C           &   On-band  &  120453  & 1200  &  1.35 \\
 IC 10 Field C           &  On+grism  &  120455  & 1200  &  1.38 \\
 IC 10 Field C           &  On+grism  &  120457  & 1200  &  1.42 \\
 IC 10 Field A           &  Off-band  &  120459  &  150  &  1.45 \\
 IC 10 Field A           &   On-band  &  120461  & 1200  &  1.48 \\
 IC 10 Field A           &  On+grism  &  120463  & 1200  &  1.54 \\
 IC 10 Field A           &  On+grism  &  120465  & 1200  &  1.62 \\
 IC 10 Field C           &  Off-band  &  120471  &  150  &  1.70 \\
 IC 10 Field C           &   On-band  &  120473  & 1200  &  1.76 \\
 IC 10 Field C           &  On+grism  &  120475  & 1200  &  1.88 \\
 IC 10 Field C           &  On+grism  &  120477  & 1200  &  2.03 \\
\hline
\end{tabular}
}
\end{minipage}
\label{tab_focas}
\end{table}

\begin{table}
\centering
\begin{minipage}{80mm}
\caption{Radial velocities and magnitudes for the detected PNe in IC10.}
\scriptsize{
\begin{tabular}{@{}lccccc@{}}
\hline
\ ID & $\alpha$&$\delta$&RV    &m$_{FOCAS}$& m$_{M03}$\\	 
       &(J2000)  &(J2000) &\kms&(5007~\AA)     &(5007~\AA)\\
\hline
01-A&0 18 44.61&59 17 47.83&-328&23.83&24.12 (1)\\
02-A&0 19 04.40&59 17 04.82&-350&23.63&24.01 (2)\\
03-A&0 19 13.64&59 14 44.55&-361&23.14&23.58 (3)\\
04-B&0 19 49.37&59 17 11.08&-371&24.83& -\\
05-B&0 19 57.71&59 20 31.42&   -&26.52& -\\
06-B$^*$&0 19 57.51&59 17 11.37&-317&24.71& \\
07-B&0 19 59.37&59 18 42.11&-361&25.00& -\\
08-B&0 20 01.15&59 16 41.55&-365&24.87& -\\
09-B&0 20 03.65&59 20 44.05&   -&24.71& -\\
10-B$^*$&0 20 03.90&59 19 26.12&-323&24.65& \\
11-B&0 20 04.20&59 16 55.30&-340&24.18& -\\
12-B&0 20 06.34&59 17 05.24&-308&25.89& -\\
13-B&0 20 06.66&59 15 22.18&-354&23.94&24.76 (4)\\
14-B&0 20 09.68&59 19 47.17&-310&23.83& -\\
15-B&0 20 09.87&59 16 34.67&-357&25.10& -\\
16-B&0 20 10.79&59 19 56.67&-344&23.92& -\\
17-B&0 20 12.75&59 19 11.53&-313&25.73& -\\
18-B$^*$&0 20 14.74&59 18 08.14&-366&25.60& \\
19-B&0 20 14.89&59 18 53.46&-320&24.50& -\\
20-B&0 20 15.82&59 14 59.75&-333&26.19& -\\
21-B&0 20 16.21&59 17 07.33&-377&26.22& -\\
22-B&0 20 17.03&59 17 58.24&-327&24.67& -\\
23-B$^*$&0 20 17.17&59 15 52.71&-342&23.64&24.49 (5)\\
24-B&0 20 17.87&59 18 08.72&-365&25.28& -\\
25-B&0 20 18.00&59 18 11.48&-333&25.69& -\\
26-B&0 20 18.14&59 18 58.17&-285&25.27& -\\
27-B&0 20 19.32&59 18 02.70&-324&22.54& -\\
28-B$^*$&0 20 19.66&59 18 13.03&-341&24.02& \\
29-B&0 20 22.28&59 16 46.59&-340&25.23& -\\
30-B&0 20 24.28&59 15 38.81&  - &25.41& -\\
31-B&0 20 27.58&59 16 53.83&  - &26.12& -\\
32-B&0 20 27.98&59 18 20.91&  - &24.78& -\\
33-C&0 20 40.65&59 17 46.32&-284&  -  & -\\
34-C&0 20 40.89&59 18 50.15&  - &22.93&23.33 (10)\\
35-C&0 20 42.95&59 16 30.90&-332&22.85&23.43 (11)\\
36-C&0 20 46.34&59 18 38.30&-372&26.43& -\\
37-C&0 20 47.40&59 19 00.05&-284&  -  & -\\
38-C&0 20 49.06&59 18 33.16&-279&24.18& -\\
39-C&0 21 09.80&59 16 59.70&-331&22.81&23.18 (15)\\
40-C&0 21 11.20&59 16 11.71&-382&23.81&24.52 (16)\\
41-C&0 21 14.40&59 18 08.17&-303&26.37& -\\
\hline                    
\end{tabular}}
{Notes --- (i) The asterisks in the 
ID column indicate the confirmed PNe studied, 
through deep medium-resolution optical spectroscopy, 
by \citet{mg09}. (ii) The last two columns 
present the magnitudes obtained in this work 
(m$_{\rm FOCAS}$, refer to the text for more details) 
and in \citet[][M03, in which the photometry 
and astrometry of a number of PN candidates in IC10 
were discussed. See their Table~1]{magrini03}. 
In the last column, the numbers in parentheses
stand for the M03's IDs. 
}
\end{minipage}
\label{tab_focas}
\end{table}

\subsection[]{(Re)identification of PNe, FOCAS slitless spectroscopy
  and Reductions}

The traditional on-band/off-band filter technique was used
for the re-detection of the known PNe 
and for the identification of new PN candidates. 
The on-band image is taken through a narrow-band filter passing the redshifted
\oiii~5007~\AA\,  nebular emission line, while the off-band image
is taken through a broader filter passing no nebular emissions.
The PNe are visible as point sources in the on-band image, but
are absent in the off-band image. A third image, taken through
the on-band filter and a grism, confirms the PN candidates. By
inserting the grism in the light path, the images of all continuum
sources are transformed into segments of width determined by
the on-band filter transmission curve. All the emission-line point
sources, such as the PNe, remain as point sources. The grism also
introduces a shift relative to the undispersed on-band image
which is a function of the wavelength of the nebular emission
line and of position on the CCD. By calibrating this shift, we
are able to measure the radial velocities for all emission-line
objects in the field, as in M\'endez et al. (2001, 2009). 
In the case of Subaru and FOCAS, the
dispersing element was an echelle grism with 175~grooves/mm 
which operates in the 4th order and gives a dispersion
of 0.5~\AA~pixel$^{-1}$, with an efficiency of 60\% (see the Subaru 
website).

Standard IRAF\footnote{IRAF is
distributed by the National Optical Astronomical Observatories,
operated by the Association of Universities for Research in
Astronomy, Inc., under contract to the National Science
Foundation} tasks were used for the basic CCD reductions 
(bias subtraction, flat field correction using twilight flats).
We refer the reader to the papers by M\'endez et al. (2001, 2009) where
a detailed description of all the steps involved in the reduction 
analysis is given.

\section[]{Kinematic Analysis}

\begin{figure} 
   \centering
   \includegraphics[width=8 truecm]{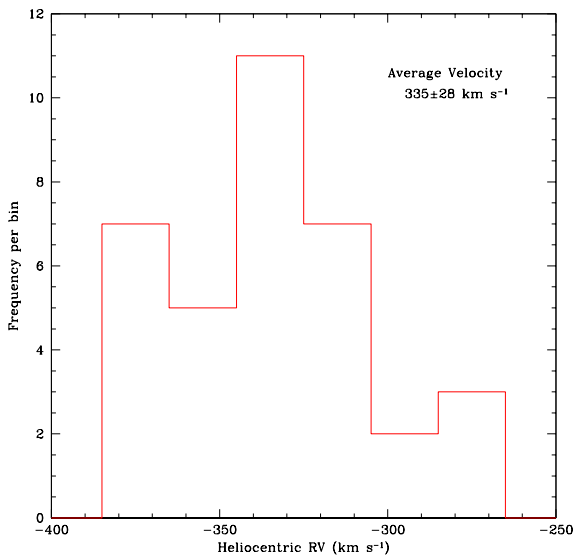} 
   \caption{Histogram of heliocentric radial velocities for 35 PN
   candidates in IC10, in bins of 20~\kms. The average velocity and
   dispersion are given in the figure.}
   \label{fig_hist}
\end{figure}

\begin{figure*} 
   \centering
   \includegraphics[width=12 truecm]{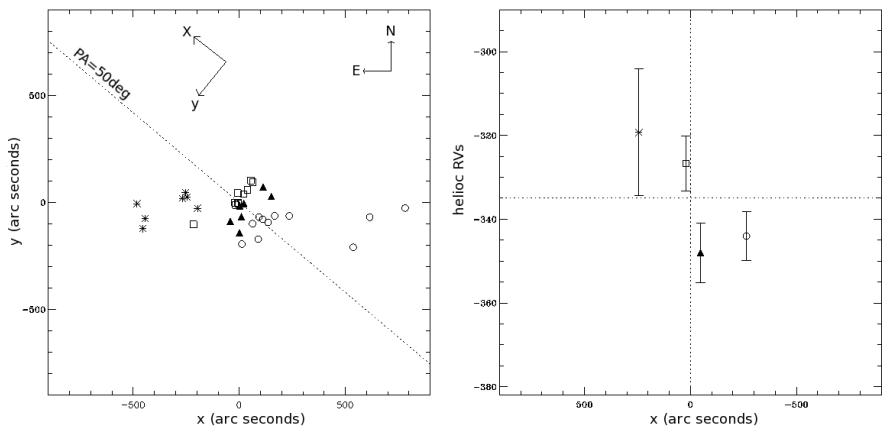} 
   \caption{$Left$: the same spatial distribution 
of PNe as in Figure~\ref{fig_FocasFields}. The
    dashed line  represents an axis of PA = 50\gr.
    To study the rotation along this axis, the PN sample was divided 
    into four bins perpendicular to it. The orientation of this axis
    is shown on the top and the PNe inside each bin are represented as: open 
    circles, filled triangles, open squares and stars. $Right$: the
    average velocity within each bin is plotted at the position of the
    average x-coordinate of the bin. The 4 bin ranges are:
$x<-100''$ (open circle), $-100''<x<0''$ (filled triangle), 
$0''<x<100''$ (open square), $x>100''$ (stars).}
   \label{fig_PA50}
\end{figure*}

\begin{figure*} 
   \centering
   \includegraphics[width=12 truecm]{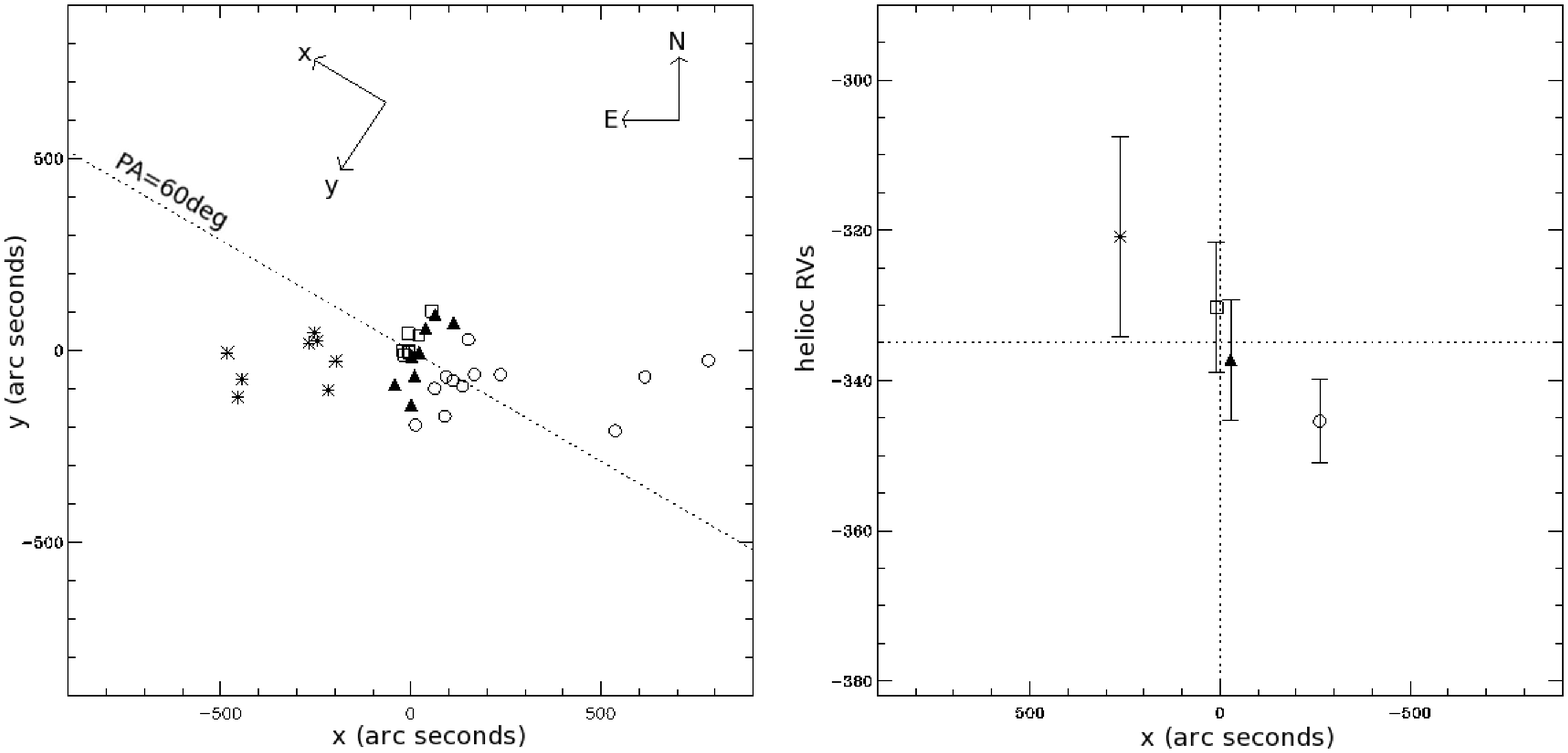} 
   \caption{As for Figure~\ref{fig_PA50}, except that the rotation is
   studied along an axis of PA=60\gr.}
   \label{fig_PA60}
\end{figure*}

\begin{figure*} 
   \centering
   \includegraphics[width=12 truecm]{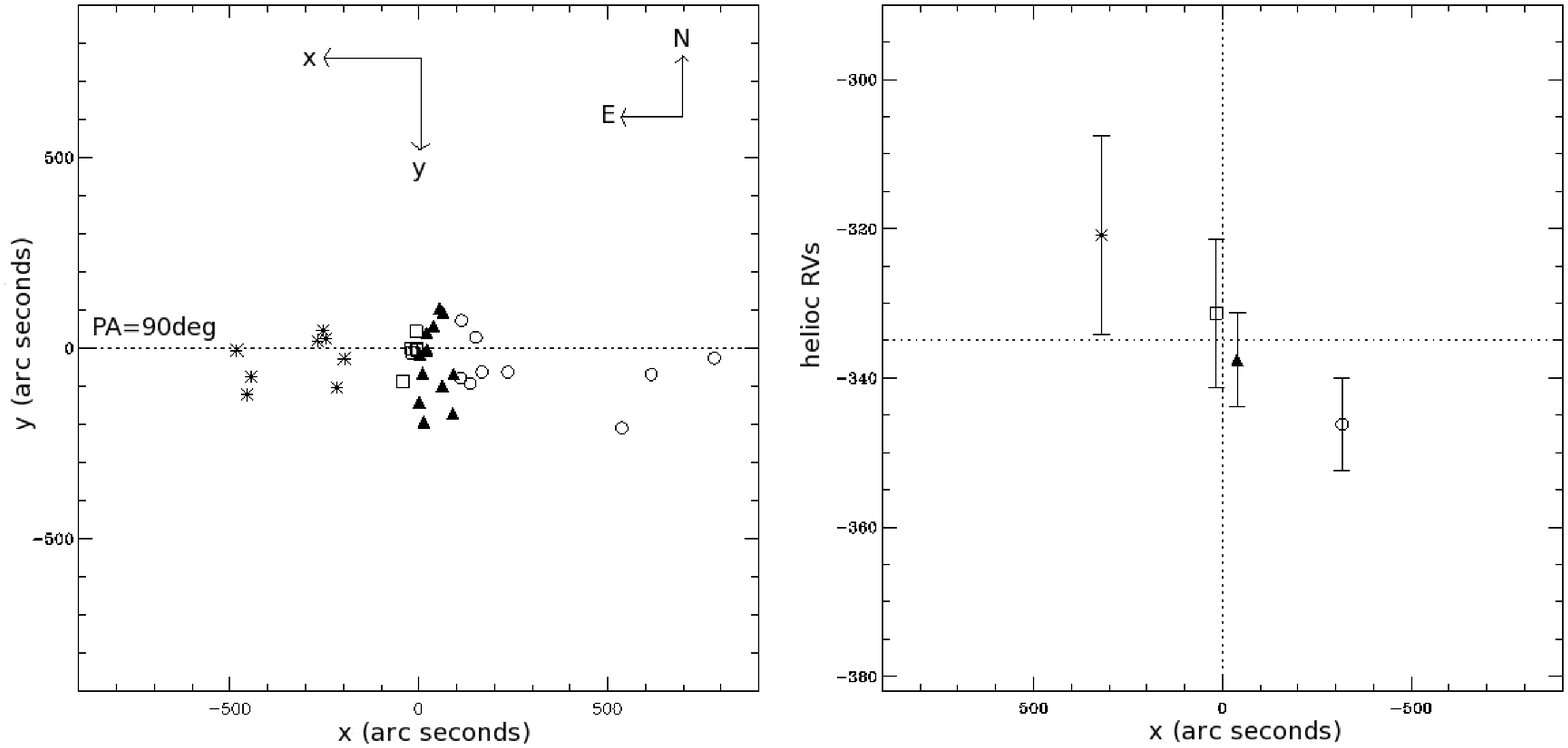} 
   \caption{As for Figure~\ref{fig_PA50}, except that the rotation is
   studied along an axis of PA=90\gr.} 
   \label{fig_PA90}
\end{figure*}

\begin{figure*} 
   \centering
   \includegraphics[width=12 truecm]{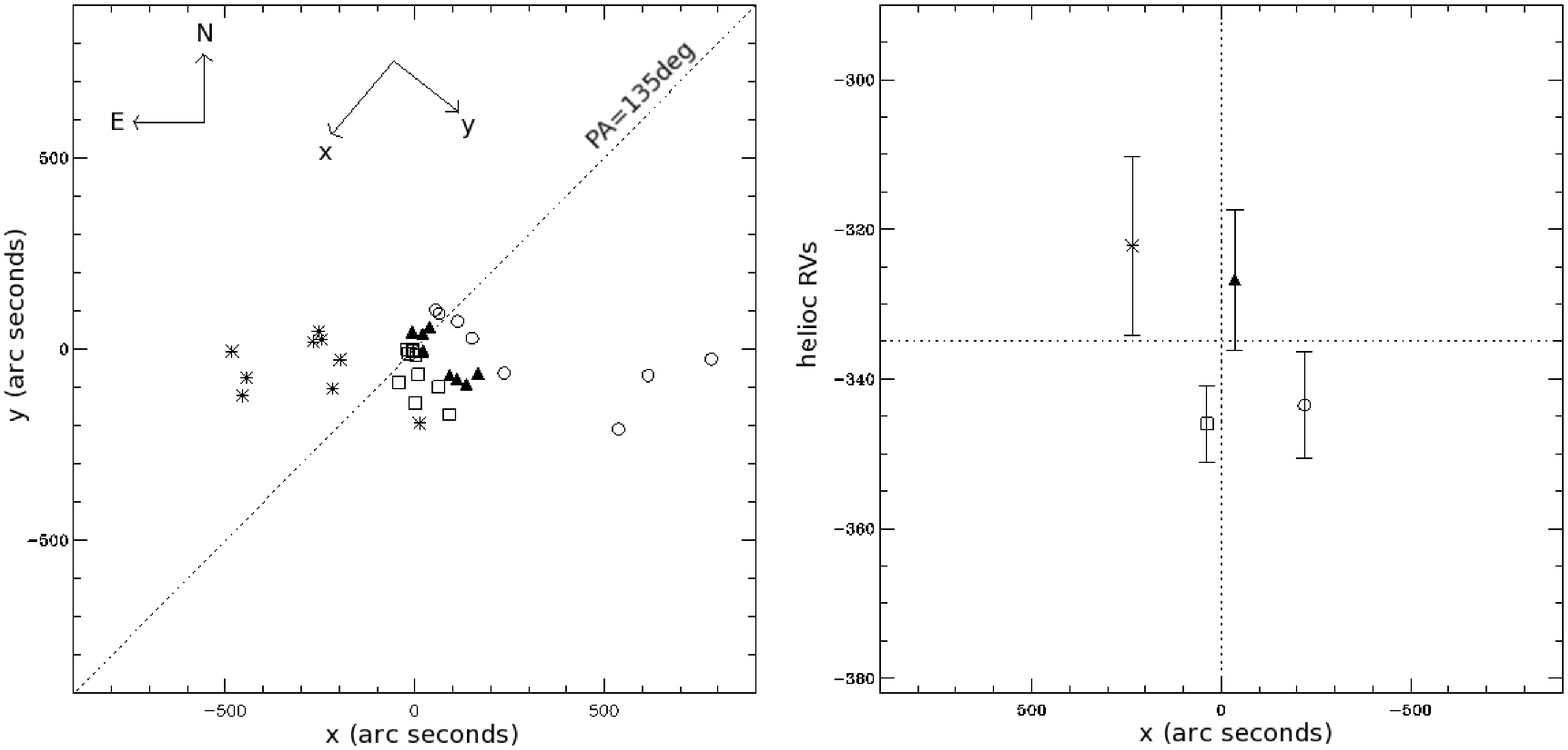} 
   \caption{As for Figure~\ref{fig_PA50}, except that the rotation is
   studied along an axis of PA=135\gr.} 
   \label{fig_PA135}
\end{figure*}

\begin{figure*}
   \centering
   \includegraphics[width=18 truecm]{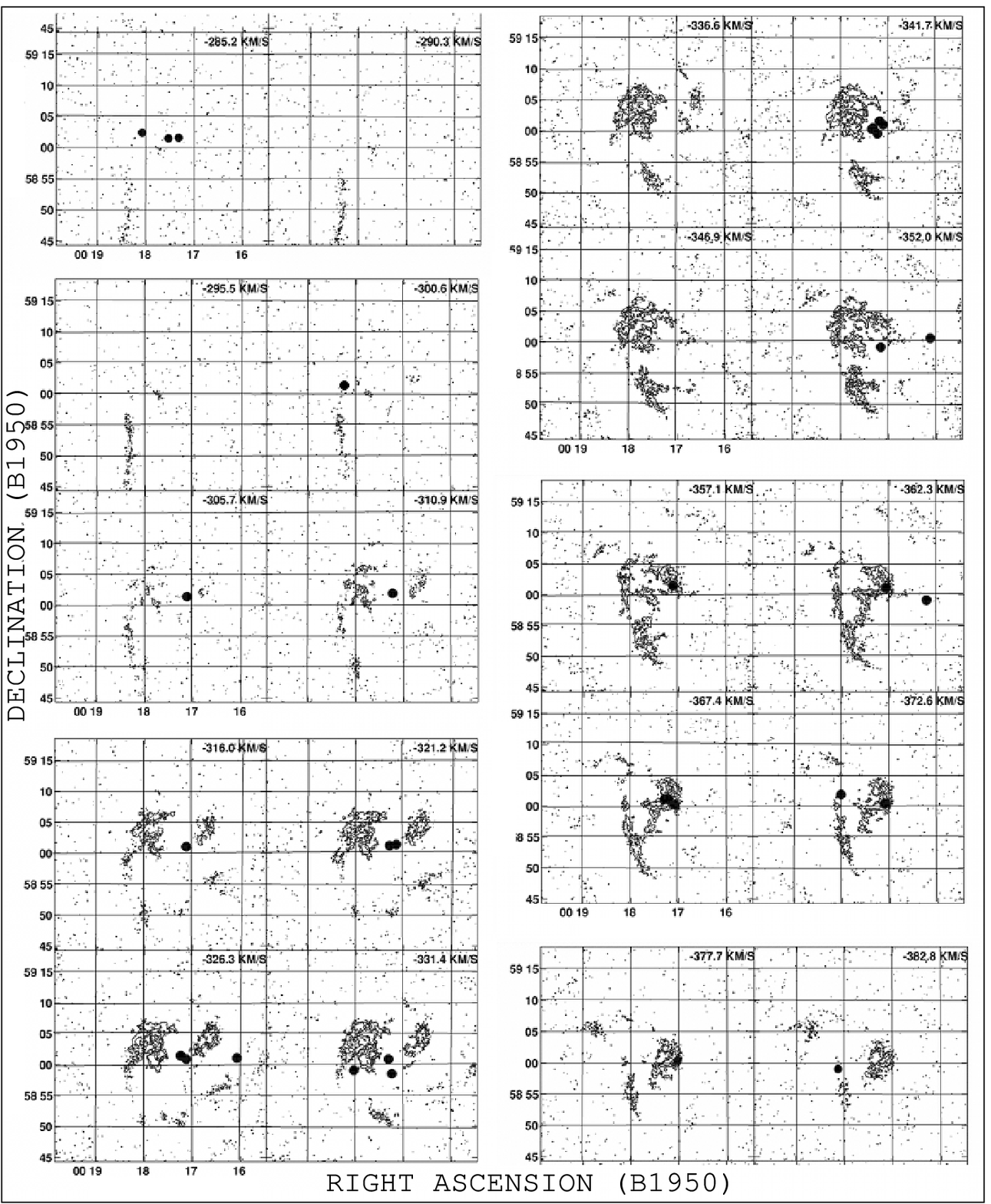} 
   \caption{The PN (filled circles) positions superimposed on the \hi\ maps
   produced by WM98. The velocities of the PNe
   in each map fall within the velocity range of the map.}
   \label{fig_maps}
\end{figure*}

In Figure~\ref{fig_FocasFields} we show the spatial distribution of the 41 PN
candidates that we have identified in IC10 with Subaru$+$FOCAS. 
As explained in Sect. 2.2, we have used the slitless technique to obtain
the heliocentric radial velocities (RVs) for the sample. Hereafter, we will
refer to our RVs simply as \lq velocities'. 
Typical images with seeing around 0.5$''$
have a point-spread function (PSF) size of 5 pixels, which
translates into a radial velocity resolution of 140~km~s$^{-1}$, 
i.e., the internal velocity field of the PN is not resolved. Assuming position
errors of 0.4 pixel, the expected uncertainty in radial velocity is
12 km s$^{-1}$.  
However, due to
background contamination, bad columns in the 
images, and the displacement produced
by the grism placing some of the sources outside of the fields, we were able to
measure the velocities of 35 objects only. The final values are listed in
Table~2. Adding quadratically the uncertainties in the velocities, which
are given by
the calibration errors ($\sim$10~$\kms$), the position errors ($\sim$10
$\kms$), and the errors from image registration ($\sim$10~$\kms$), we
get an overall error of about 17~$\kms$. Assuming that the
spectrograph deformations and guiding errors have a marginal
contribution, we estimate the total uncertainty in the velocities
measured with Subaru to be at most 20~$\kms$. 

We show in Figure~\ref{fig_hist} a histogram of the velocities 
measured for our targets. There is a clear velocity peak at
$-$335\kms, with a dispersion of $\pm$28~\kms. Within the
uncertainties, the average PN velocity is in good agreement 
with the NED heliocentric radial velocity of $-$348\kms\ for IC10. 
We note from the velocity histogram 
that none of the objects is apart from the others
in terms of velocity distribution. 
This finding suggests that the objects are true
members of IC10. 

We then obtained the galactocentric coordinates $x$ and $y$ 
for the objects. For that purpose we have used a 
central position of $\alpha$(J2000)=0h 20m 17.3s
and $\delta$(J2000)=59\gr18\arcmin14\arcsec\ for IC10. As
IC10 is an irregular galaxy, it is difficult to define its dynamical axis. 
\citet{wilcots98} argue that even though somewhat
uncertain the position angle 
(PA) value might range between 50 and 90 degrees (see. e.g.,  
their Figure~5). For this reason, we chose to investigate
any sign of rotation coming from the PNe along axes with different
PA, slightly extending the WM98's PA range. 
Along each of those axes we defined four subsamples
and calculated the average velocity for each one. We note, however,
that the subsamples were not equally spaced 
in $x$ because we want to ensure a minimum number
of PNe per subsample. 
Figures 3 to 6 show the binned velocities of the PNe
along axes with PA of 50\gr, 60\gr, 90\gr and 135\gr, respectively. 
We selected these angles because PA=135\gr corresponds to the major
axis of the IC 10 starlight distribution (as can be seen in Figure 1),
while PA=60\gr gives the direction along which the neutral gas is 
rotating, according to SS89 and WM98.

Figures~\ref{fig_PA50}, \ref{fig_PA60} and \ref{fig_PA90} suggest
some rotation along the axis with PA= 50\gr, 60\gr and 90\gr,
while figure~\ref{fig_PA135} shows that the evidence of rotation
along the optical PA=135\gr axis is less significant. We conclude that
the PNe (and therefore the stellar population) appear to be rotating 
like the neutral gas.

Let us make a more detailed comparison.
In WM98, the authors produced velocity maps at different 
locations in the sky, each map corresponding to a velocity channel. 
In Figure~\ref{fig_maps} 
the position of our PNe with velocities falling within the range of 
the \citet{wilcots98} map channels are  
superimposed to the corresponding map. We note that in most maps PNe 
are found to coincide 
with the gas, further suggesting that the PN kinematics follows 
the \hi\ one. Note that WM98 concluded, 
using the velocity maps we just discussed, 
that their \hi\ velocities followed the rotation curve obtained 
previously by SS89. So we can also compare our PN velocities to the 
plots in SS89. In Figure 8 we have taken the position-velocity diagram
from SS89 (their Fig. 10) and we have superimposed our PN velocities.
This diagram was constructed with an $x$ coordinate measured along 
PA=60\gr, following the observed neutral gas rotation. Along this
direction, the H I velocities are very well behaved, as is most 
clearly seen in SS89's Fig 8. Turning to our objects, in spite of
the high PN velocity dispersion, it is immediately
apparent that the PNe are rotating in the same direction, with 
less negative velocities in the upper left. There is a good overlap with 
the neutral gas velocities, although several PN velocities are less
negative. The full interpretation of this figure is not straightforward,
because, although the $x$ coordinates are the same, the $y$ coordinates 
can be quite different. Consider for example the three PNe in Field A
(see Figure 1). Their $x$ coordinates are the most negative in our sample
($< -500$), and they might be expected to lie at the most negative 
velocities (they appear at the extreme right of Figure 8); but their average 
velocity is $-340$ km s$^{-1}$. However, they should not be compared with the
neutral gas velocity distribution shown in Figure 8, because they are 
in a different part of the sky, to the northwest. In fact, there is a blob 
of \hi\ close to these Field A PNe, seen in SS89's Fig. 9, and it has a 
velocity of about $-330$ km s$^{-1}$, which fits those few PN velocities 
quite well, perhaps by chance.

In summary, leaving aside some discrepancies, which probably but not 
necessarily 
indicate a tendency toward more negative neutral gas velocity than PN 
velocities, in Figs. 3 to 8 we have been able to show that most of the 
PN velocities follow closely the \hi\ kinematics 
pattern derived by SS89 and WM98. This illustrates 
the regular, solid-body rotation-like nature of the velocity 
field (given by the \hi\ as well the PNe) in 
IC10's disk.

\begin{figure} 
   \centering
   \includegraphics[width=8.5 truecm]{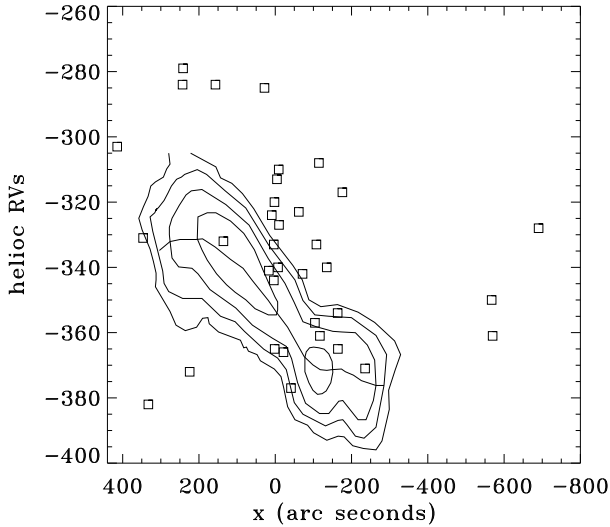} 
   \caption{The position-velocity diagram from Figure 10 of SS89,
showing the kinematics of neutral gas along PA=60\gr. The squares
indicate our PN velocities as a function of the corresponding $x$ 
coordinate, calculated along an axis with PA=60\gr, as in Figure 4.} 
   \label{fig_HI+PNe}
\end{figure}

\section[]{PN photometry}

A detailed description of the PN photometric procedure is provided by
\citet{teodorescu10}.
The results of our photometry for 39 of the 41 objects are listed in 
Table~2. Two objects (PN 33 and 37) were 
too close to bright foreground stars, which 
precluded accurate photometry. The \oiii~5007~\AA\ fluxes measured 
through the on-band filter are
traditionally expressed in magnitudes m(5007), using the definition
introduced by \citet{jacoby89}.
\begin{equation}
m(5007) = -2.5 \log\ I(5007) -13.74.
\end{equation}
For the flux calibration, we adopted the standard star G138-31 \citep{oke90},
which has a monochromatic flux at 5028~\AA\ 
(the redshifted wavelength of its \oiii~5007\AA\, 
emission due to the PN radial velocity), of  
1.44 $\times$ 10$^{-15}$ ergs cm$^{-2}$ s$^{-1}$ ~\AA$^{-1}$. 

If we compare our photometry with that by \citet{magrini03} for the
objects in common, we find that our values are systematically brighter,
by about 0.4 mag.
We attribute this to the fact that the nights of \citet{magrini03} 
were not of photometric quality, and thus our new values presented in
Table~2 are preferred.

\subsection{Foreground extinction}

Being located at a low galactic latitude (l=119.0\gr, b=-3.3\gr), 
IC10 is affected by significant foreground extinction. As previously
stated, we find  in the literature different values for this parameter. 
\citet{schlegel98} find $E(B-V)=1.527$, while studies based either on the 
tip of the red giant branch (TRGB;  \citealt{sakai99}; \citealt{sanna08}) 
or the Cepheids \citep{wilson96} give a range between 0.6 to 1.1 mag. We 
have adopted $E(B-V)=0.77$, following \citet{ri01}, whose 
extinction value is preferred 
because of the very convincing work they did, by plotting the 
reddening as a function of 
the \hi\ column density in their Figure~3.    

\subsection{The distance determination from the PN luminosity function}

\begin{figure} 
   \includegraphics[width=9 truecm]{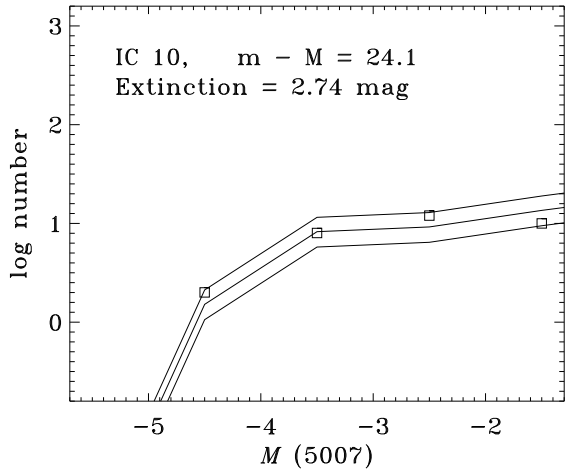} 
   \caption{Observed \oiii~5007 PNLF of IC10 (squares), with the
     sample of 35 PNe binned into 1-mag intervals. The apparent
     magnitudes m(5007) have been transformed into absolute magnitudes
   M(5007) by adopting an extinction correction of 2.74 mag and a
   distance modulus of 24.1. The three lines are PNLF simulations
   (M\'endez and Soffner 1997) for 3 different total PN population
   sizes: 100, 150 and 200.}
   \label{fig_pnlfIC10b10dm241}
\end{figure}

Having measured the apparent magnitudes
m(5007) of the 35 PNe, the PN luminosity function (PNLF) was built,
using 1-mag bins, and compared with simulated PNLFs like those used
by \citet{ms97} to fit the observed PNLF of M31. The
comparison is shown in 
Figure~\ref{fig_pnlfIC10b10dm241}. The absolute magnitudes M(5007) that
produce the best fit to the simulated PNLF were calculated using an
extinction correction A($\lambda$), at 5007~\AA, of 3.56$E(B-V)$ 
\citep{ccm89}, and a distance modulus $(m - M)$ = 24.1, which is 
equivalent to 660~kpc. 

The simulated PNLFs plotted in Figure~\ref{fig_pnlfIC10b10dm241} 
are binned, like the
observed one, into 1-mag intervals and have a maximum final mass of
0.63~\sm\ and sample sizes between 100 and 200 PNe (see \citealt{ms97}; the 
``sample size'' is the total number of PNe, detected
or not, that exist in the surveyed area). We selected a rather large bin 
size of 1 mag to compensate for the rather small number of PNe, and 
reduce the statistical noise in the PNLF. We estimated an error of 0.2
mag from the goodness of the fit at different distance moduli. To
obtain the total error estimate, we have to combine the possible
systematic and random errors. The systematic error is the same as in
\citet{jcf90}, i.e., 0.13 mag, including the possible error in
the distance to M31, in the modeling of the PNLF and in the
foreground extinction. The random contributions are given by 0.2 mag
from the fit to the PNLF, 0.05 mag from the photometric zero point,
and 0.05 mag from the filter calibration. Combining all these errors
quadratically, we estimate that the total error bar for the PNLF
distance modulus is $\pm$ 0.25 mag. A PNLF was obtained by \citet{kniazev08}, 
based on \citet{magrini03} magnitudes. Given the uncertainties in both
determinations, \citet{kniazev08}'s distance modulus
(24.30$^{+0.18}_{-0.10}$) is in good agreement with ours. 
Note that we have adopted brighter magnitudes than \citet{magrini03}.
If we correct \citet{kniazev08}'s distance modulus by the same amount,
their distance modulus decreases to 23.9.

By studying the TRGB in IC10 \citet{sakai99} 
found a distance modulus of 24.1$\pm$0.2~mag for Population I
and 23.5$\pm$0.2~mag for Population II, while \citet{sanna08} estimated
a distance modulus of 24.57$\pm$0.15 mag. From the observations of the
Cepheids, \citet{wilson96} deduced a distance modulus of
24.57$\pm$0.21 mag. Note that there may be some amount of internal
extinction in IC10, in addition to the foreground extinction. 
For the sake of a complete knowledge of the IC10 properties, 
although not significantly 
affecting our conclusions about the kinematic pattern of the galaxy, 
it would be important to know 
how much extra extinction to expect. Obviously we cannot go further 
on this discussion without 
deep individual spectra for all the PNe we found. Such a work, clearly, lies 
outside of the scope of the present paper.  Had we  
increased the extinction correction to 3.5 mag, equivalent to $E(B-V)$=1.0 
\citep{demers04}, our new distance modulus would become 23.3. 
Such a large correction is unlikely, because many of the objects we 
discovered are in fact in the outskirts of IC10. We give these numbers to 
emphasize that the value of the extinction correction is by far the most 
important source of uncertainty in the PNLF distance determination for 
IC10.

\subsection{PN formation rate}

Once the sample size is known, we can calculate the specific PN
formation rate $\dot{\xi}$ in units of PNe yr$^{-1}$ $L_{\odot}$$^{-1}$: 

\begin{equation}
n_{PN} = \dot{\xi} L_{T} t_{PN},
\end{equation}
where $n_{PN}$ is the sample size, $L_{T}$ is the total bolometric
luminosity of the sampled population, expressed in $L_{\odot}$, 
and $t_{PN}$ is the lifetime of a PN, for which we have adopted 30,000~yr 
in the PNLF simulations. We have the B band mag, $B_{T}$=8.68, 
and $B-V$=0.53,  
from \citet{ri01}, and a bolometric correction of $-0.8$ mag \citep{bu06} 
from which we obtain an extinction corrected apparent
bolometric magnitude 7.35. Using the distance modulus we obtained, 
of 24.1, and a
solar $M_{bol}$ = 4.72 we calculate the total luminosity of IC 10, 
$L_{T}$ = 3.9 $\times$ 10$^{8}$ $L_{\odot}$.  The sampled luminosity
is 90\% of the total luminosity, i.e. 3.5 $\times$ 10$^{8}$
$L_{\odot}$. Adopting $n_{PN}$ = 150 (see subsection above), we get  
$\dot{\xi}$ = (14.28 $\pm$  5) $\times$ 10$^{-12}$. Transforming our
value of $\dot{\xi}$ into a specific PN density --or $\alpha$ ratio, 
defined as the number of PNe to the total galaxy luminosity ratio--, 
as log ($\alpha)$ = $-6.37$, in agreement with IC10'starburst quality 
(\citealt{bu06} obtained $-6.59$).

\section{Discussions}

Our motivation to study the PN kinematics of IC 10
is based on the fact that some PN candidates were previously found in this
galaxy \citep{magrini03}, and, moreover, located at large
galactocentric distances (around 3~kpc, see Figure~\ref{fig_FocasFields}) 
from its centre, PN
1 to 3 in Table 2. The latter can represent a trace of past tidal
interactions and/or the connection of the old-intermediate stellar
population with the \hi\ envelope extending 7 times farther than the
optical diameter of the galaxy (\citealt{huctmeier79}; \citealt{hsm81};
see also Meatheringham et al. 1988 for the LCM). So far, no stellar 
kinematics (of individual objects) was available (for any kind of stars) 
in IC10, thus the link between stars and gas was missing, and the 
hypothesis that the \hi\ envelope was a later acquisition, rotating 
in the opposite
direction from that of the inner parts of the galaxy (WM98), needed 
supporting evidence.

Previous to the present work there have been no kinematic 
studies of the
PN population in IC10. Although IC10 is a dwarf galaxy, where not many 
PNe can be expected to be found (because of its low luminosity),
we were able to collect a sample of 35 PNe. A small sample in absolute 
terms, but the largest detected to date in this kind of galaxy,
with the only exception of the Magellanic clouds (see Table
1 of Magrini 2006,  with the number of PNe per LG galaxy, and also
\citealt{mg09} and \citealt{gon12} for an update). It is
particularly relevant that PNe form a population with intermediate
age between the \hi\ and massive young stars and the old population of red
giant stars, and therefore a study of their kinematics should cast light
on the problem of the different rotation patterns of the inner disk and its
surrounding outer gas (see, e.g.,  SS89).

The presence of the PN population in the 
outskirts of IC10 implies that there was 
active star formation in the outer envelope 
in the past, several billions of years ago.
Unfortunately, without a deep spectroscopic 
study of the PNe at the outskirts of the galaxy 
we cannot better date the star formation episode 
that gave birth to their central stars. Such 
a study does exist, but only for the central 
5.5$\times$5.5~arcmin$^2$ region of IC10 \citep{mg09}. 
Following the latter authors, the 
spectroscopically observed PNe, all with low N/O ratio and/or 
with very low or absent He II, are, to a first 
approximation, \lq\lq old stars". And, 
thus, they could be born in a limited period 
of time, 7 to 11~Gyr ago, that is, during the first 
half of the age of the Universe. 
This is in 
agreement with other indications coming from 
photometric studies of the stellar population in 
the outskirts of IC 10, e.g., those by \citet{sanna09} who identified 
stars belonging to different star formation episodes 
(massive young stars ($\sim$100 Myr), old 
stars ($\sim$13 Gyr), and intermediate age stars 
($\sim$0.2-7~Gyr)). In addition, \citet{sanna10} 
found a  significant decrease of the young stellar 
population when moving from the center toward 
the outermost regions of IC10. They  detected 
several samples  of old red giant stars (well fitted 
with isochrones of 13 Gyr) up to radial distances 
of 18-23~arcmin from the galactic center, and an  
old star excess at least up to 34-42~arcmin from the 
center. Similar results were also obtained by 
\citet{tikh09}, who detect an old stellar population 
coincident with the \hi\ envelope.
Thus we can suppose that the external PNe are 
associated with the old and intermediate-age 
stellar populations, since no young population 
is detected in the outskirts. 
Note that we did not detect any \hii\ 
region in the FOCAS field A, which suggests that, presently, 
there is no ongoing star formation 
in the outskirts of IC10. Moreover, being a starburst
galaxy, with a current star formation 
rate ranging from 0.02 to 0.71~M$_{\odot}$~yr$^{-1}$ 
(see \citealt{mateo98} and \citealt{yin10} 
compilations), IC10 is undergoing an intense and very recent burst 
of star formation \citep{yin10}, probably 
involving only its more central parts. 
\citet{yin10} detailed
modelling of IC10 chemical evolution showed, 
as others previously suspected \citep{mg09},  
that IC10 was probably formed by means of a slow gas 
accretion process with a long infall timescale of $\sim$8~Gyr. 
This time scale is particularly significant 
because it corresponds to the epoch of the formation of the 
PN progenitors in the central regions of IC10. 
If  the progenitors of the PNe located in 
the outskirts of IC10 had the same age of the central ones, 
they might be the signature of an epoch of  
particularly  active star formation also in the outskirts of IC10. 
We might suppose that this important episode 
of star formation was driven by  the first  
encounter of the \hi\ gas cloud with the IC10 proto-galaxy. 
The conclusions by Sanna et al. (2010) are in very good 
agreement with our findings,  
supporting  the hypothesis that the \hi\ cloud is associated with IC10
and that it was hosting star formation in the past.

Concluding, the above discussion is in agreement 
with the numerical simulation by 
\citet{dimatteo} who investigated the enhancement of star formation efficiency 
in galaxy interactions and mergers using numerical simulations, and 
considering different types of encounters, both direct and retrograde. 
They found that  retrograde encounters produce  
a larger star formation efficiency  than direct encounters. 
They explain this behaviour by the fact that in 
retrograde encounters tides are less efficient, allowing most of the initial 
gas to stay in the galaxy, rather than to be 
driven outwards. This great reservoir of gas can furnish the fuel 
for an intense burst of star formation in the merging phase.

\section{Summary}

The study described in this paper was motivated by the fact that a previous 
search for PNe in IC10 found some PNe located at 
the outskirts of this starburst galaxy (up to 3~kpc from 
the centre). Their location by
itself suggests that star formation was active 
in the past at the very outer regions
of the galaxy. Thus, we took advantage of the deep imaging and 
spectroscopic capabilities provided by FOCAS@Subaru to add a number of 29 
new objects to the previously known 
(\citealt{magrini03}; \citealt{kniazev08}; \citealt{mg09}) 
PNe of IC10, simultaneously measuring the 
radial velocity of most of the galaxy's PNe.

The average radial velocity of 35 of our PNe 
is -~335$\pm$28~\kms, which is, within the errors, 
the same as the recession velocity of the 
galaxy (-348~\kms; NED). The PN radial velocities
were compared with the \hi\ gas velocities in 
order to examine whether the kinematic connection with 
the underlying stellar population of IC10 could 
represent the trace of past tidal
interactions in the Local Group. We found that 
the \hi\ and PN population share the 
same rotation pattern, as evidenced by the 
rotation curves along any PA between 50 and 90\gr, following 
\citet[and our Figures 3-5 and 8]{shostak89}, 
or the velocity maps of \citet[shown in Figure 7]{wilcots98}. 
Moreover, considering that the PNe located 
at high galactocentric distances in IC10 have roughly the same age 
as those in the inner regions, around 8~Gyr 
(\citealt{mg09}; \citealt{yin10}), the first encounter of the 
\hi\ gas with the galaxy can be constrained to have occurred at that time. 

Finally, our deep \oiii~5007~\AA\ FOCAS images 
were used to build up the planetary
nebulae luminosity function. From the latter 
we derived a IC10 distance modulus 
of 24.1$\pm$0.25, which corresponds to D = 660~kpc. 
This distance is in very good agreement with
other recent IC10 distance determinations 
(\citealt[24.1$\pm$0.2;]{sakai99} and 
\citealt[24.30$^{+0.18}_{-0.10}$]{kniazev08}).

\section{Acknowledgments}
DRG kindly acknowledges the UCL Astrophysics Group, for their hospitality, 
where part of this work was done. 
AAB acknowledges support from ARC (Super Science Fellowship, FS110200016).
LM is supported through the ASI-INAF grant ``HeViCS: the Herschel Virgo 
Cluster Survey" I/009/10/0. 
This work was also supported by the National Science Foundation (USA) under 
grant 0807522.

\bsp

\label{lastpage}

\end{document}